\documentclass{article}

\usepackage{arxiv}

\usepackage[utf8]{inputenc} 
\usepackage[T1]{fontenc}    
\usepackage{hyperref}       
\usepackage{url}            
\usepackage{booktabs}       
\usepackage{amsfonts}       
\usepackage{nicefrac}       
\usepackage{microtype}      
\usepackage{graphicx}
\usepackage{doi}

\usepackage{bm}
\usepackage{subcaption}
\usepackage{ulem}

\usepackage{amsmath}
\usepackage[sorting=nyt, style=numeric]{biblatex}
\title{A comparison of methods for estimating\\ the average treatment effect on the treated\\ for externally controlled trials}

\author{Huan Wang\thanks{Contributed equally.} \\
	Division of Biometrics IX\\
	OB/OTS/CDER, FDA,\\
	Silver Spring, MD 20993 \\
	\texttt{huan.wang@fda.hhs.gov} \\
\And
    Fei Wu\footnotemark[1] \\
	Division of Biometrics IX\\
	OB/OTS/CDER, FDA,\\
	Silver Spring, MD 20993 \\
	\texttt{fei.wu1@fda.hhs.gov} \\
\And
    Yeh-Fong Chen\thanks{Corresponding author.} \\
	Division of Biometrics IX\\
	OB/OTS/CDER, FDA,\\
	Silver Spring, MD 20993 \\
	\texttt{yehfong.chen@fda.hhs.gov} \\
}

\renewcommand{\headeright}{}
\renewcommand{\shorttitle}{Estimating Average Treatment Effects on Treated}

\hypersetup{
pdftitle={A comparison of methods for estimating the average treatment effect on the treated for externally controlled trials},
}

\begin{document}
\maketitle

\begin{abstract}
While randomized trials may be the gold standard for evaluating the effectiveness of the treatment intervention, in some special circumstances, single-arm clinical trials utilizing external control may be considered. The causal treatment effect of interest for single-arm trials is usually the average treatment effect on the treated (ATT) rather than the average treatment effect (ATE). Although methods have been developed to estimate the ATT, the selection and use of these methods require a thorough comparison and in-depth understanding of the advantages and disadvantages of these methods. In this study, we conducted simulations under different identifiability assumptions to compare the performance metrics (e.g., bias, standard deviation (SD), mean squared error (MSE), type I error rate) for a variety of  methods, including the regression model, propensity score matching (PSM), Mahalanobis distance matching (MDM), coarsened exact matching, inverse probability weighting, augmented inverse  probability weighting (AIPW), AIPW with SuperLearner, and targeted maximum likelihood estimator (TMLE) with SuperLearner. 

Our simulation results demonstrate that the doubly robust methods in general have smaller biases than other methods. In terms of SD, nonmatching methods in general have smaller SDs than matching-based methods. The performance of MSE is a trade-off between the bias and SD, and no method consistently performs better in term of MSE. The identifiability assumptions are critical to the models' performance: Violation of the positivity assumption can lead to a significant inflation of type I errors in some methods; violation of the unconfoundedness assumption can lead to a large bias for all methods. 

According to the simulation results, under most scenarios we examined, PSM and MDM methods perform best overall in terms of type I error control. However, they in general have worse performance in the estimation accuracy compared to doubly robust methods given that the identifiability assumptions are not severely violated.
\end{abstract}


\section{Introduction}
Randomized clinical trials (RCTs) are usually considered the gold standard for demonstrating the efficacy and safety of an intervention. However, for rare diseases it is difficult to recruit a large number of patients, and for severe diseases it is unethical to assign patients to placebo. Single-arm trials utilizing an external real-world data (RWD) control could be an appealing option. In addition to the feasibility and ethical advantages, single-arm trials with external RWD controls may save time and money and as a result have been receiving increasing attention.

To draw inferences about the effect of a treatment, a target causal effect needs to be defined. For the single-arm trials with external RWD controls, the causal treatment effect of interest could be the average treatment effect on the treated (ATT) (i.e., the average treatment effect for the population receiving the treatment compared to what would have happened if they had not received treatment) if the treatment effect among the enrolled population is of more interest~\cite{heckman1985alternative,heckman1997matching}. 
This is particularly the case if there is heterogeneity between patients participating and those not participating in the trial due to eligibility criteria.  For example, if the experimental drug is a second-line treatment, the study drug's effect cannot be considered for naive patients who are not eligible to participate in the trial.

The ATT can also be of interest if there is heterogeneity in the baseline characteristics between patients participating and those who do not. E.g., patients who participate in the trial are older than patients who do not. 
In this study, we focus on estimating the causal effect of ATT.

Many well-established methods have been proposed to estimate ATT, including the coarsened exact matching (CEM), propensity score matching (PSM), Mahalanobis distance matching (MDM), inverse propensity weighting (IPW), augmented inverse propensity weighting (AIPW), targeted maximum likelihood estimator (TMLE), etc. Attempts have been made to compare the performance of these models in terms of estimating ATT. For example, Abdia et al. compared the performance of the PSM, regression model, stratification model, and IPW~\cite{abdia2017propensity}; Chatton et al. compared the performance of the g-computation method, IPW, PSM, and TMLE with different covariates sets~\cite{chatton2020g}. However, the choice of the model remains unclear and controversial in the actual design stage of single-arm clinical trials with RWD controls.

In this study, we examined the performance of 10 methods for causal inference, including the regression model, PSM, MDM, CEM, IPW, AIPW, AIPW with SuperLearner, and TMLE with SuperLearner via simulations. The metrics of performance used for comparison included bias, standard deviation (SD), mean squared error (MSE), and type I error. With our simulation results, practical advice on the use of these methods in single-arm clinical trials with RWD controls is provided. 

This paper is organized as follows. The introduction section provides background information on single-arm trials with external RWD controls and presents the motivation behind this study. The Methods section revisits the definitions of ATT and identifiability assumptions and provides a description of the models and simulation settings utilized in this study. The Results section presents the findings from various scenarios. In the Discussion section, we analyze the experimental results and provide recommendations for planning single-arm trials with external controls. Finally, the Conclusion section summarizes the study's key findings, interpretation, and implications.

\section{Methods}
\subsection{Average treatment effect on the treated}
In a real-world study, a causal effect of a treatment can be defined in terms of the comparison between potential outcomes (also known as the counterfactual outcomes) of a population who would receive the treatment and potential outcomes of the same population who would not receive the treatment~\cite{hernan2020causal}. The comparison is drawn from a population level instead of an individual level because only one of the potential outcomes can be observed for each patient~\cite{abadie2005semiparametric}. In a nonrandomized, single-arm clinical trial with an external RWD control, the population of interest is usually only the patients receiving the treatment~\cite{fang2020statistical}. Correspondingly, the causal effect of interest is the ATT.

For patient with index $i$, let $Z_i$ denote the patient's treatment assignment with a value of 1 for treated and 0 for not treated, $\bm{X}_i$ vector of the patient's covariates, $Y_i$ the patient's observed outcome, $Y_i(1)$ and $Y_i(0)$ the patient's potential outcomes corresponding to $Z_i=1$ and $Z_i=0$, respectively. Then the ATT of our interest in terms of the difference scale $\tau_{ATT}$ rather than others (e.g., ratio), can be formally defined as
$$\tau_{ATT}\equiv E[Y_i(1)-Y_i(0)|Z_i=1].$$
\subsection{Identifiability assumptions}
An average causal effect such as ATT is not always estimable. Hern{\'a}n and Robins summarized three assumptions needed to estimate an average causal effect: consistency, positivity, and conditional exchangeability~\cite{hernan2020causal}. The definitions of these assumptions are as follows.
\begin{enumerate}
\item Consistency: $Y_i=Y_i(1)$ if $Z_i=1$ and $Y_i=Y_i(0)$ if $Z_i=0$. 
\item Positivity: $0<P(Z_i=1|\boldsymbol{X_i=x_i})<1$ for any $\bm{X}_i$ with $P(\boldsymbol{X_i=x_i})>0$. 
\item Conditional exchangeability (unconfoundedness): $Y_i(0), Y_i(1)  \perp\kern-5pt\perp Z_i||\bm{X}_i.$
\end{enumerate}

The consistency assumption means that a subject’s
potential outcome under his or her observed exposure history is equal to the subject’s observed outcome. The consistency assumption requires that the treatment should be sufficiently well-defined, contingent upon agreement among experts based on the available substantive knowledge~\cite{hernan2020causal}.

\begin{figure}
\includegraphics[width=1\linewidth]{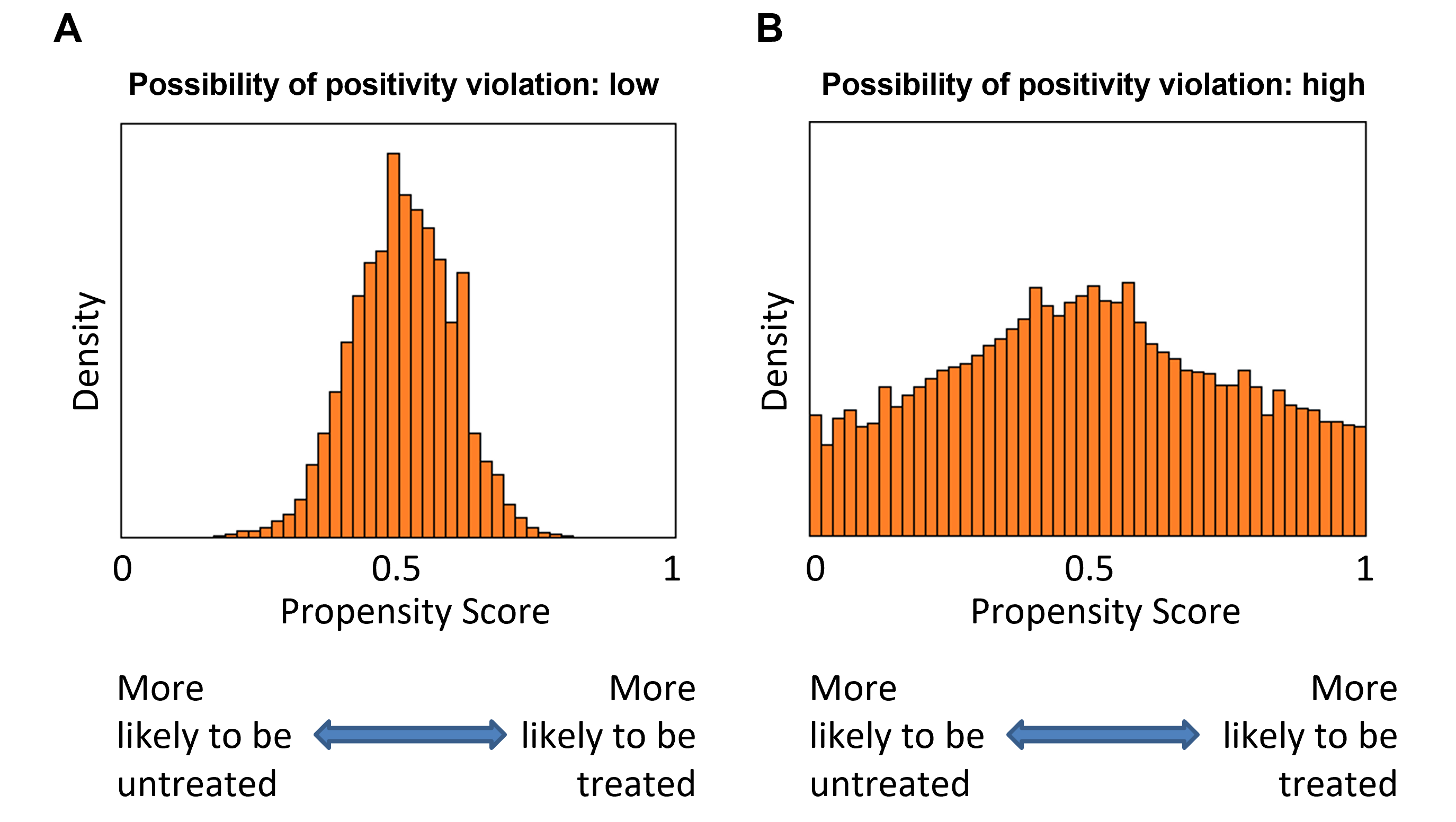}
		\caption{Distributions of PS values in study populations characterized by low (A) and high (B) potential for violating the positivity assumption. In these populations, subjects with PS values close to 0 or 1 are more prone to violating the positivity assumption.}	\label{fig:positivity_assumption}
\end{figure}

The positivity assumption implies that all subjects have a non-zero probability of receiving (or not receiving) the treatment. Positivity violations can be divided into two categories: 1) theoretical violation that the probability of being treated is 0 or 1, e.g., certain patients are restricted from receiving the treatment or control, and 2) near or practical violation in the sampling due to high/low exposure prevalence or small sample size~\cite{leger2022causal}. One method for assessing the positivity assumption is via the propensity score (PS), which is the probability of receiving the treatment conditional on covariate values, i.e., $PS(\bm{X}_i)=P(Z_i=1|\bm{X}_i)$~\cite{zhu2021core}. In particular, we can study the distribution of the PS for the study population and explore whether there are subjects with PS values near 0 or 1 (Figure~\ref{fig:positivity_assumption}). If many PS values are distributed around 0 or 1, then the positivity assumption should not hold; on the other hand, if no PS values are distributed around 0 or 1, the positivity assumption can be considered reasonable.

The conditional exchangeability assumption, also known as the unconfoundedness assumption, suggests that all variables influencing treatment selection and outcomes are measured and accessible. Violation of this assumption occurs when unmeasured confounders are not adequately accounted for. The unconfoundedness assumption is inherently untestable. However, many sensitivity analyses have been developed to indirectly assess the unconfoundedness assumption, i.e., assessing the bias of the causal effect estimate by assuming unmeasured confounding~\cite{rosenbaum1983assessing,imbens2003sensitivity,ding2016sensitivity,vanderweele2017sensitivity}. 

\subsection{Methods for causal inference}\label{section_methods}

\subsubsection{Linear regression model}
In general, a regression model is not a sound model for estimating ATT. However, if the true outcome model can be properly fitted by a regression model, the regression model may estimate ATT well. In particular, if 1) the identifiability assumptions are met and 2) the outcome model is correctly specified, the ATT can be estimated through the regression model. However, in practice, it is challenging to specify the correct outcome model. In this study, we use the linear regression (LR) that includes the specified covariates $\bm{X}_i$ and the treatment variable $Z_i$ without interaction terms as a simple reference model. The ATT is estimated through the coefficient of the treatment variable $Z_i$.

\subsubsection{Matching}
A fundamental problem of causal inference is that we can only observe one potential outcome from each observation. However, the information from the potential outcome that we do not observe is needed in the estimation of the causal effect. An intuitive approach is to estimate the unobserved potential outcome with a similar one via matching. An underlying assumption about the matching is that observations similar on their covariate values are also similar on the potential outcomes~\cite{king2019propensity}. For estimating ATT using matching, we can assume that the expectation of $Y_i(0)$ and $Y_i(0)^M$ are the same where $Y_i(0)^M$ is the sample in the control group that matches $Y_i(1)$ on $\bm{X}_i$, i.e., 
\begin{equation}
E[Y_i(0)|\bm{X}_i]= E[Y_i(0)^M|\bm{X}_i].\label{equation: match on X}
\end{equation}
Then by the unconfoundedness assumption, ATT can be expressed as 
\begin{align*}
 \tau_{ATT}
 =&E[Y_i(1)-Y_i(0)|Z_i=1]\\
 =&E_{\bm{X}_i|Z_i=1} \{E[Y_i(1)|\bm{X}_i]-E[Y_i(0)|\bm{X}_i]\} \\
 =& E_{\bm{X}_i|Z_i=1} \{E[Y_i(1)|\bm{X}_i]-E[Y_i(0)^M|\bm{X}_i]\}\\
 = & E[Y_i(1)-Y_i(0)^M|Z_i=1]  
\end{align*}
and estimated by
$$\frac{1}{N_1}\sum_{i\in{[Z=1}]}[Y_i(1)-Y_i(0)^M],$$ where $N_1$ is the sample size of the treatment group.
    
\subsubsection{Coarsened exact matching}

In practice, exact matching of covariates may not be feasible when there are continuous covariates or when there are categorical covariates with many levels. One approach to address this issue is the CEM method, which is a form of exact matching based on the subclasses of covariates~\cite{iacus2011multivariate}. It involves two steps: 1) coarsening each covariate into several ``bins,'' i.e., subclasses, and 2) performing exact matching on those bins. The unmatched units are discarded.

For CEM, the choice of the number of bins is crucial to model performance. Domain knowledge is required to determine the appropriate number of bin for each variable~\cite{iacus2011multivariate}. In this study, we arbitrarily choose 2 (denoted by CEM2), a smaller number of bin, and 5 (denoted by CEM5), a reasonably larger one, for all covariates to study the performance of the CEM model with different numbers of bins. For each number of bins, the boundaries of the bins were set to be evenly spaced between the maximum and minimum values of the covariate~\cite{ho2018package}.

\subsubsection{Mahalanobis distance matching}

An alternative way to loosen the exact match requirement is by using MDM so that similar subjects are matched. MDM pairs subjects that are close based on the Mahalanobis distance. The Mahalanobis distance between Patients $u$ and $v$, denoted by $d_{u,v}$, is defined as 
	$$
	d_{u,v}=\sqrt{
		(\bm{X}_u-\bm{X}_v)^\top\bm{\Sigma}^{-1}(\bm{X}_u-\bm{X}_v)
	},
	$$
where $\bm{\Sigma}$ is the covariance matrix of the covariates.

\subsubsection{Propensity score matching}
Matching can also be conducted on the PS of the relevant covariates instead of on the covariates themselves. Rosenbaum and Rubin proved that PS is a balancing score and that potential outcomes are independent of treatment conditional on a balancing score if the unconfoundedness assumption holds~\cite{rosenbaum1983central}. Then the assumption of unconfoundedness for the given covariates can be relaxed to be unconfoundedness for the given PS values, i.e., 
$$Y_i(0), Y_i(1)  \perp\kern-5pt\perp Z_i \ |  \ {X_i} \ \Rightarrow Y_i(0), Y_i(1)  \perp\kern-5pt\perp Z_i \ | \  PS(\bm{X}_i).$$

In addition, according to equation~(\ref{equation: match on X}) and the law of total expectation, we have
\begin{align}
E[Y_i(0)|PS(\bm{X}_i)]
=&E\{E[Y_i(0)|\bm{X}_i]|PS(\bm{X}_i)\} \nonumber \\
= & E\{E[Y_i(0)^{PSM}|\bm{X}_i]|PS(\bm{X}_i)\} \nonumber \\ 
= &E[Y_i^{PSM}(0)|PS(\bm{X}_i)], \label{equation: match on PS}
\end{align}where $Y_i(0)^{PSM}$ is the sample in the control group that matches $Y_i(1)$ on $PS(\bm{X}_i)$.
Then, by  the assumption of the unconfoundedness given the PS and equation~(\ref{equation: match on PS}), ATT can be obtained by
{\small
 \begin{align*}
    \tau_{ATT}=  
   &E[Y_i(1)-Y_i(0)|Z_i=1] \\
    =&E_{PS(\bm{X}_i)|Z_i=1} \{E[Y_i(1)|PS(\bm{X}_i)]-E[Y_i(0)|PS(\bm{X}_i)]\}\\
    =&E_{PS(\bm{X}_i)|Z_i=1} \{E[Y_i(1)|PS(\bm{X}_i)]-E[Y_i(0)^{PSM}|PS(\bm{X}_i)]\} \\
    =&E[Y_i(1)-Y_i(0)^{PSM}|Z_i=1]
 \end{align*}
 }
 and estimated by
    $$\frac{1}{N_1}\sum_{i\in{[Z=1}]}[Y_i(1)-Y_i(0)^{PSM}].$$

For a comprehensive introduction of matching-based methods, including PSM, refer to the overview by Stuart~\cite{stuart2010matching}.
	
 \subsubsection{Inverse probability weighting}

The IPW is another PS-based method that can be used to estimate the ATT.  The IPW uses the PS to balance the characteristics of covariates in the treated and control groups by weighting each subject in the analysis by the inverse probability of receiving his/her actual exposure~\cite{chesnaye2022introduction}.  In this study, we used a generalized version of the IPW estimator, which has the following expression:

\begin{align*}
   \frac{\sum_{i=1}^n W_iZ_iY_i}{\sum_{i=1}^n W_iZ_i} -\frac{\sum_{i=1}^n W_i(1-Z_i)Y_i}{\sum_{i=1}^n W_i(1-Z_i)} \\
\end{align*}
where $W_i=\frac{\widehat{PS}(\bm{X}_i)}{Z_i\widehat{PS}(\bm{X}_i)+(1-Z_i)(1-\widehat{PS}(\bm{X}_i))}$ ~\cite{li2018balancing,mao2019propensity}.
This IPW estimator is a consistent estimator of ATT when the identifiability assumptions hold and $\widehat{PS}(\bm{X}_i)$ is a consistent estimator of the PS.

\subsubsection{AIPW}
The AIPW method combines the IPW method and an outcome estimation model. It is a doubly robust estimator in the sense that it is consistent if either the treatment assignment model or the potential outcome model is correctly specified, but not necessarily both~\cite{robins1995analysis}. In this study, we used the following AIPW estimator for ATT estimation:
\begin{align*}
    \frac{\sum_{i=1}^n \widehat{PS}(\bm{X}_i))(\widehat{Q}_1(\bm{X}_i))-\widehat{Q}_0(\bm{X}_i))}{\sum_{i=1}^n \widehat{PS}(\bm{X}_i)} 
    +\frac{\sum_{i=1}^n W_iZ_i(Y_i-\widehat{Q}_1(\bm{X}_i))}{\sum_{i=1}^n W_iZ_i} 
    -\frac{\sum_{i=1}^n W_i(1-Z_i)(Y_i-\widehat{Q}_0(\bm{X}_i))}{\sum_{i=1}^n W_i(1-Z_i)} 
\end{align*}
where 
$W_i=\frac{\widehat{PS}(\bm{X}_i)}{Z_i\widehat{PS}(\bm{X}_i)+(1-Z_i)(1-\widehat{PS}(\bm{X}_i))}$ and $\widehat{Q}_1(\bm{X}_i)$ and $\widehat{Q}_0(\bm{X}_i)$ denote the estimators for $E[Y_i|\bm{X}_i, Z_i=1)]$ and $E[Y_i|\bm{X}_i, Z_i=0)]$, respectively ~\cite{mao2019propensity}.

\subsubsection{Targeted maximum likelihood estimator}

The TMLE, introduced by van der Laan and Rubin, is an alternative doubly robust estimation method that is maximum-likelihood–based and includes a targeting step that optimizes the bias-variance tradeoff for the parameter of interest~\cite{van2006targeted,schuler2017targeted}.

The application of TMLE for estimating ATT involves the following steps~\cite{van2011targeted,fang2020two}:

\begin{enumerate}
\item Obtain an initial estimate of the expected outcome $\widehat{E}^0 [Y_i|Z_i, \bm{X}_i]$ by fitting an outcome model. 

\item Obtain an initial estimate of the PS, i.e.,  $\widehat{PS}^0(Z_i|\bm{X}_i)$, by fitting a treatment assignment model.

\item Apply the targeted learning theory to update the initial estimates of $\widehat{E}^0 \left[Y_i | Z_i, \bm{X}_i\right]$ 
and $\widehat{PS}^0 \left(Z_i |\bm{X}_i\right)$ 
to $\widehat{E}^\ast \left[Y_i | Z_i, \bm{X}_i\right]$ 
and $\widehat{PS}^\ast \left(Z_i |\bm{X}_i\right),$ respectively. These updates are performed such that they solve the efficient score equation: 
$$\sum_{i=1}^{n}D\left(\widehat{E}^\ast \left[Y_i | Z_i, \bm{X}_i\right],{\widehat{PS}}^\ast \left(Z_i | \bm{X}_i\right),{\widehat{P}^\ast}_{ \bm{X}_i|Z_i=1}\right)=0$$
where $D$ is defined as
{\scriptsize
\begin{align*}
   D\left(E\left[Y_i | Z_i,  \bm{X}_i\right], PS\left(Z_i | \bm{X}_i \right), P_{ \bm{X}_i|Z_i=1}\ \right)
=&\left[\frac{I\left(Z_i=1\right)}{P\left(Z_i=1\right)}-\frac{I\left(Z_i=0\right)PS(Z_i=1| \bm{X}_i)}{P\left(Z_i=1\right)PS(Z_i=0| \bm{X}_i)}\right]
(Y_i-E\left[Y_i | Z_i, \bm{X}_i\right])\\
&+\frac{I\left(Z_i=1\right)}{P\left(Z_i=1\right)}\left(E\left[Y_i | Z_i=1,  \bm{X}_i\right]-E\left[Y_i | Z_i=0,  \bm{X}_i\right]-\tau_{ATT}\right)
\end{align*}
}

and $\widehat{P}^\ast_{\bm{X}_i|Z_i=1}$ is the empirical estimator of $P_{ \bm{X}_i|Z_i=1}$.

\item The TMLE estimator for the ATT is then obtained as:
{\scriptsize
\begin{align*} \widehat{\tau}_{ATT}^\ast\ =\int{\left({ \widehat{E}}^\ast \left[Y_i | Z_i=1,  \bm{X}_i\right]-{ \widehat{E}}^\ast \left[Y_i | Z_i=0,  \bm{X}_i\right]\right)d{{ \widehat{P}}^\ast}_{ \bm{X}_i|Z_i=1}}.
\end{align*}
}

\end{enumerate}

In this study, we implemented TMLE by using the SuperLearner for the estimation of both propensity scores and outcomes~\cite{gruber2006package}. The model is denoted as TMLE\_SL. In addition, we implemented a model, denoted as AIPW\_SL, on the basis of the AIPW method and the SuperLearner's estimation of the propensity scores and outcomes.

\subsection{Simulation}
The performance of the methods described in Section~\ref{section_methods} for estimating ATT was compared by using Monte Carlo simulations. To evaluate the methods in practical situations where the identifiability assumptions may be violated, we considered the following three scenarios:

\begin{itemize}
	\item \textit{Scenario 1:} No violation of identifiability assumptions,  
	\item \textit{Scenario 2:} Violation of the positivity assumption,
	\item \textit{Scenario 3:} Violation of the unconfoundedness assumption.
\end{itemize}

Note that we did not consider the scenario where the consistency assumption was violated because the consistency could be ensured by proper study design and conduct. 

In the simulation, the patient's treatment status was generated by the treatment selection model based on the covariates, while the patient's outcome was generated by the outcome model based on the covariates and treatment status.  

In \textit{Scenarios 1} and \textit{2}, we assumed three independent covariates $X_1 - X_3$ that followed the standard normal distribution.  $X_1$ was a confounder that affected both the treatment selection and the outcome, and $X_2$ and $X_3$ affected the treatment selection and the outcome, respectively. 
In \textit{Scenario 3}, an additional confounder $X_4$ was assumed to follow the standard normal distribution and to be independent of $X_1 - X_3$.

The treatment selection models for corresponding scenarios used in our simulations are listed in Table \ref{trt models}. For each subject, treatment status, denoted by $Z_i$, was generated from a Bernoulli distribution with a probability of success of $p_{i,treat}$. We considered five different prevalence rates of treatment, i.e., expectations of  $p_{i,treat}$ with respect to the distributions of covariates: 0.05, 0.10, 0.20, 0.33, and 0.50.  The intercept $\alpha_{0,treat}$ in each treatment selection model was determined numerically to obtain the desired treatment prevalence.

\begin{table}[h!]
	\scalebox{0.9}{
	\begin{tabular}{ll}
		\hline\\
		\textit{Scenario 1}&   
		$\mathrm{logit}(p_{i,treat})
		= \alpha_{0,treat} + 0.1X_{1,i} + 0.1 X_{2,i}+  0.05X_{1,i}^2 + 0.02 X_{2,i}^2 
		+ 0.02 X_{1,i}X_{2,i}$\\
		\\
		\textit{Scenario 2}& 
		$\mathrm{logit}(p_{i,treat})
		=\alpha_{0,treat} + 1.25X_{1,i} + X_{2,i}+  0.5X_{1,i}^2 + 0.5 X_{2,i}^2 
		+ 0.75 X_{1,i}X_{2,i}$   \\
		\\
		\textit{Scenario 3}& 
		$\mathrm{logit}(p_{i,treat})
		=		\alpha_{0,treat} + 0.1X_{1,i} + 0.1 X_{2,i}+  0.05X_{1,i}^2 + 0.02 X_{2,i}^2 
		+ 0.02 X_{1,i}X_{2,i} + 0.05X_{4,i} + 0.02X_{4,i}^2$   \\
		\\
		\hline
	\end{tabular}}
\caption{\label{trt models}Treatment selection models}
\end{table}

For each scenario, we considered three settings of outcome models:
\begin{itemize}
	\item \textit{Setting 1:} Linear outcome model.
	\item \textit{Setting 2:} Nonlinear outcome model.
	\item \textit{Setting 3:} Outcome model with the interaction term between the treatment variable and the covariate.
\end{itemize}

The outcome models used in our simulations to reflect these settings are listed in Table \ref{out models}, where we assumed $\epsilon_i\sim N(0,2)$. 

\begin{table}[h!]
	\scalebox{1}{
		\begin{tabular}{ll}
			\hline\\
			\textit{Setting 1}&  
   			$Y_i =\begin{cases}
                        Z_i + 1.5X_{1,i} + 0.75 X_{3,i} + \epsilon_i, & \text{\ \ \ \ \ \ \ \ \ \ \ \ \ \ (\textit{Scenarios 1\&2})}\ \\
                        Z_i + 1.5X_{1,i} + 0.75 X_{3,i} + 5 X_{4,i} + \epsilon_i, & \text{\ \ \ \ \ \ \ \ \ \ \ \ \ \ (\textit{Scenario 3})}\
                        \end{cases}$ \\
			\textit{Setting 2}& 
			$Y_i =\begin{cases}
                        Z_i + 1.5X_{1,i} + 0.75 X_{3,i} + 1.75 X_{1,i}^2  + \epsilon_i, & \text{(\textit{Scenarios 1\&2})}\ \\
                        Z_i + 1.5X_{1,i} + 0.75 X_{3,i} + 1.75 X_{1,i}^2 + 5 X_{4,i} + \epsilon_i, & \text{\  (\textit{Scenario 3})}\
                        \end{cases}$ \\
			\textit{Setting 3}&
             $Y_i =\begin{cases}
                        Z_i + 1.5X_{1,i} + 0.75 X_{3,i} + 1.5X_{1,i}Z_i + \epsilon_i, & \text{(\textit{Scenarios 1\&2})}\ \\
                        Z_i + 1.5X_{1,i} + 0.75 X_{3,i} + 1.5X_{1,i}Z_i +  5 X_{4,i} + \epsilon_i, & \text{(\textit{Scenario 3})}\
                        \end{cases}$ \\
                        \\
                        \hline
	\end{tabular}}
	\caption{\label{out models}Outcome models}
\end{table}

We performed 200 simulations for each prevalence rate of treatment under each combination of treatment selection model and outcome generation model. The steps of simulations are described below.
\begin{enumerate}
    \item Generated $N$ sets of covariates from independent standard normal distributions, assuming that each set was associated with one patient. The sample size $N$ was determined by making the expected number of treated patients equal to 50, e.g., $N=1000$ if the prevalence rate of treatment was 0.05, and $N=500$ if the prevalence rate of treatment was 0.10.
    \item For each set of the covariates associated with a patient, compute $p_{i,treat}$ by using the corresponding treatment selection model. The patient's treatment status $Z_i$ was then generated via the Bernoulli distribution with a success probability of $p_{i,treat}$. The patient's outcome $Y_i$ was generated by using the corresponding outcome model.
    \item Estimated the ATT by using the methods described in Section~\ref{section_methods}. For each method, we included only covariates $X_1 - X_3$ for estimation. We used the logistic regression based on $X_1 - X_3$ to estimate the PS for PSM, PSM\_1:2, IPW, and AIPW. And we used the LR based on $X_1 - X_3$ to estimate the outcomes for AIPW. 
    \item Repeated the above process 200 times.
    \item Evaluated the performance of each method through following metrics:
\begin{itemize}
	\item The bias in estimating ATTs, i.e., the average difference between the 200 estimated ATT and the true ATT.  Note that the true ATTs in \textit{Settings 1} and \textit{2} are 1. For \textit{Setting 3}, it is difficult to calculate the true ATT analytically and we approximated the true ATT by using a large number of simulated samples (i.e., $10^7$) from the treatment models described in Table~\ref{trt models}.
	\item The empirical SD of the estimated ATTs, i.e., the SD of 200 estimated ATTs.
	\item The average theoretical SD of the estimated ATTs, i.e., the average of 200 model-based SDs, each generated directly by the model in each simulation.
	\item The MSE in estimating ATTs, i.e., the MSE of 200 estimated ATTs.
 \item The type I error rate, i.e., the average rate of rejecting the null hypothesis of $\text{ATT}=0$ versus the alternative hypothesis of  $\text{ATT} \neq 0$ using a two-sided test at the 0.05 significance level  by making $Z_i=0$ in the outcome models described in Table~\ref{out models}.
 \end{itemize}
\end{enumerate}

Note that we set the expected number of treated subjects as 50 to reflect the practical situation that the feasible number of patients who can be enrolled in the study requiring the external control is often limited.

In this study,  the nearest neighbor matching without replacement algorithm with a caliper of 0.2 was used in forming matched pairs for PSM, PSM\_1:2, and MDM to improve the balance of the matching and reduce the MSEs of the estimates~\cite{austin2011optimal,austin2014comparison}; a trimming threshold of 0.05 was used for estimated the PS for IPW and AIPW for better variance estimation~\cite{austin2022bootstrap};  a truncation threshold of  $5/\left(\sqrt{N} \text{ln}(N)\right)$ was used for estimated the PS for AIPW\_SL and TMLE\_SL to minimize both bias and MSE of the estimates~\cite{gruber2022data}.

For the CEM, MDM, and PSM implemented in this study, the average theoretical SD of the ATT were estimated through the paired t-test between two groups in matched samples to account for the dependence induced by the matching~\cite{austin2009type}. For PSM\_1:2, the estimator-based SD was also estimated through the paired t-test in matched samples where each treated sample is compared to the average of the two matched control samples.

For this study,  all statistical analyses were conducted in R (Version 4.2.2) using the following libraries:
MatchIt (Version 4.5.5), PSweight (Version 1.1.8), tmle (Version 2.0.0), and their respective dependencies.

\section{Results}

We report results for the three different scenarios: 1) no violation of the identifiability assumptions, 2) violation of the positivity assumption, and 3) violation of the unconfoundedness assumption separately in the following subsections.

\subsection{\textit{Scenario 1}: no violation of the identifiability assumptions}
The distribution of the PS values for \textit{Scenario 1} assuming no violation of the positivity or the unconfoundedness is shown in Figure S1 in the \textit{Supplementary Material}. As shown in the figure, the distributions are bounded away from 0 and 1, and thus the positivity assumption can be considered reasonable. The bias, empirical SD (hereafter referred to as SD),  average theoretical SD, MSE, and type I error rate for the 10 methods in \textit{Scenario 1} under three different settings are shown in Figures S2-S4 in the \textit{Supplementary Material}, respectively. 

In \textit{Setting 1} with linear effect of covariates in the true outcome model, all methods have very small biases. The matching-based methods (i.e., CEM2, CEM5, PSM, PSM\_1:2, and MDM) have relatively larger SDs than the other methods for most prevalence rates of treatment, which results in greater MSEs of these methods. Among the matching-based methods, PSM\_1:2 has a relatively smaller SD than PSM when the prevalence rate of treatment is less than 0.5, while CEM2 has a relatively smaller SD than CEM5. 
The IPW, AIPW, AIPW\_SL, and TMLE\_SL models  in general  have similar performance in terms SD and MSE. Under the null treatment effect, the type I error rates of all methods are generally controlled at 0.05.

In \textit{Setting 2} with nonlinear effect of covariates in the true outcome model, the biases of most models increased compared with \textit{Setting 1}. The CEM2  in general presents the largest bias while the CEM5 presents a relatively small bias, indicating that CEM model with two bins is too coarse to catch the nonlinear effect.
The MDM and LR models  in general yield the second and third largest biases, respectively. 
The AIPW\_SL and TMLE\_SL have the smallest biases and SDs and hence the smallest MSEs.  The PSM in general has the largest SD and therefore a relatively large MSE.
Under the null treatment effect, many methods in \textit{Setting 2} have larger type I error rates than in \textit{Setting 1} (e.g., CEM2, MDM, and LR),  due to their increased biases in \textit{Setting 2}.

In \textit{Setting 3} with the interaction between the treatment variable and the covariate in the true outcome model, most methods yield very small biases. 
In terms of SD, the matching-based methods in general have relatively larger SDs than the other methods. The MSE pattern is similar to the SD pattern. 
Under the null treatment effect, the outcome models for \textit{Settings 1} and \textit{3} are the same, therefore, the findings regarding the type I error are also the same for these two settings.

\subsection{\textit{Scenario 2}: violation of the positivity assumption}
The distribution of the PS values for \textit{Scenario 2} assuming violation of the positivity or the unconfoundedness is shown in Figure S1 in the \textit{Supplementary Material}. As shown in the figure, a large proportion of subjects' PS values are distributed at or close to 0 or 1 and thus the positivity assumption is considered violated. The bias, empirical SD,  average theoretical SD, MSE, and type I error rates for the 10 methods in \textit{Scenario 2} under three different settings are shown in Figures S5-S7 in the \textit{Supplementary Material}, respectively. 

In \textit{Setting 1} with linear effect in the true outcome models, CEM2 and IPW in general have noticeably larger biases compared to other methods. The increased bias of CEM2 in \textit{Scenario 2} compared to \textit{Scenario 1} is attributed to the impact of violating the positivity assumption. Under \textit{Scenario 2}, many subjects cannot be matched exactly but are still matched when using two bins, thus causing a larger bias. 
The IPW also produces a relatively larger bias under \textit{Scenario 2}.
The SD pattern in \textit{Scenario 2} differs from that in \textit{Scenario 1} in the following ways: 1) The SDs of IPW, AIPW, and TMLE\_SL have a noticeable increase, and 2) the average theoretical SD of TMLE\_SL is much smaller than the empirical SD. 
In terms of MSE, LR and AIPW\_SL are the best two performing models. 
Under the null treatment effect, CEM2, IPW, and TMLE\_SL experience remarkable type I error inflation. According to Figure S5 in the \textit{Supplementary Material}, the type I error inflation of the CEM2 or IPW is mainly due to the large bias, while the type I error inflation of the TMLE\_SL is mainly due to its underestimation of the SD. In contrast, PSM and PSM\_1:2 perform best in terms of type I error control.

In \textit{Setting 2} with nonlinear effect in the true outcome models, in addition to the CEM2 and IPW, which are already substantially biased in \textit{Setting 1}, LR and AIPW are also largely biased. The bias of CEM5 is relatively large when the prevalence rate of treatment is low and relatively small when the prevalence rate of treatment is high. The biases of PSM and PSM\_1:2 are the smallest. The SDs of most models in \textit{Setting 2} increase compared to \textit{Setting 1}, especially for IPW and LR. The AIPW\_SL has a relatively small bias and the smallest SD, resulting the smallest MSE.   Under the null treatment effect, LR, CEM2, IPW,  AIPW, and CEM5 show severe type I error inflation, mainly due to their large biases (Figure S6 in the \textit{Supplementary Material}); TMLE\_SL also shows severe type I error inflation, mainly due to its underestimation of the SD (Figure S6 in the \textit{Supplementary Material}).   AIPW\_SL has relatively mild type I error inflation, while PSM, PSM\_1:2, and MDM have the best type I error control.

In \textit{Setting 3} with the interaction between the treatment variable and the covariate in the true outcome model, the IPW, LR, and matching-based methods in general have  relatively large biases (Figure S7 in the \textit{Supplementary Material}). The doubly robust methods (i.e., AIPW, AIPW\_SL, and TMLE\_SL) in general have relatively small biases. Among the doubly robust methods, AIPW and TMLE\_SL have relatively large SDs and hence relatively large MSEs. The MSE of AIPW\_SL is the smallest, due to its relatively small bias and SD.  The findings for \textit{Setting 3} regarding the type I error inflation are the same as those for \textit{Setting 1} (Figure S7 in the \textit{Supplementary Material}).

\subsection{\textit{Scenario 3}: violation of the unconfoundedness assumption}
For all methods and three settings of the outcome model, violation of the unconfoundedness assumption in \textit{Scenario 3} leads to a certain amount of bias (or a larger bias if the method is already biased) compared to \textit{Scenario 1} where this assumption is not violated (Figures S8-S10 in the \textit{Supplementary Material}). The biases induced by violation of unconfoundedness assumption are all positive under our simulation settings. We conducted post-hoc analyses by varying the coefficient values of the unmeasured confounder in the three settings and found that the sign and magnitude of the induced bias are mainly determined by the coefficient value of the unmeasured confounder.

The pattern of SD is similar between \textit{Scenario 1} and \textit{Scenario 3}. The MSEs observed in \textit{Scenario 3} are greater than those observed in \textit{Scenario 1}. Under the null hypothesis, the type I error rates in \textit{Scenario 3} generally increase compared to \textit{Scenario 1} due to the larger biases.

\section{Discussion}
\subsection{Performance of estimation}
\subsubsection{Matching-Based Methods}
Among matching-based methods, our simulation results show that the PSM model in general results in a small bias when the positivity assumption is held. When the positivity assumption is violated and when there is interaction between the treatment variable and the covariate (i.e., \textit{Scenario 2} \textit{Setting 3}), a mild bias can be induced. This is because under the violation of the positivity assumption, the integration of $E[Y_i(1)|PS(\bm{X}_i),Z_i=1]-E[Y_i(0)|PS(\bm{X}_i),Z_i=0]$ in equation~(\ref{equation: match on PS}) is  carried out in the overlapped range of the PS distributions for the two groups and hence the causal effect of interest has been deviated from the ATT. Similar issues occur in the other matching-based methods that we studied. Of note, the bias of the PSM can still be small in cases with positivity violation but without interaction between the treatment variable and the covariate, e.g., \textit{Scenario 2} \textit{Settings 1} or \textit{2}, because in equation~(\ref{equation: match on PS}), $E[Y_i(1)|PS(\bm{X}_i),Z_i=1]-E[Y_i(0)|PS(\bm{X}_i),Z_i=0]\\=E[E[Y_i(1)|\bm{X}_i,Z_i=1]-E[Y_i(0)|\bm{X}_i,Z_i=0]|PS(\bm{X}_i)]$
 is a constant, and its integration over $PS(\bm{X}_i)|Z_i=1$ is the same as its integration in the overlapped range of the PS distribution. 

The bias of PSM\_1:2 was comparable to that of PSM. However, the MSE of PSM\_1:2 is generally smaller than PSM due to a smaller SD. 

When considering the choice of matching ratio, our simulation indicates that 1 to 2 matching provides a reasonable balance between bias reduction and precision for estimating the ATT in most of our settings with the prevalence rate of treatment $\leq 0.33$. This finding aligns with the simulation results of Austin~\cite{austin2010statistical} for observational studies, which demonstrated that matching one or two untreated subjects to each treated subject generally provides good performance for ATT estimation. For scenarios where the control group is substantially larger than the treatment group and high-quality matches are available, using 1 to r matching with r $>$ 2 may be considered to improve efficiency, though maintaining match quality is needed.  

The performance of MDM was in general similar to PSM. The MDM has a relatively large bias in \textit{Scenario 1} \textit{Setting 2} because we used Mohalanobis distance to match only first-order terms and the second-order terms were not well matched. 
Through simulations, we found that if we also consider the second order terms (i.e., treat $X_1^2$, $X_2^2$, and $X_3^2$ as three additional covariates to be matched) in the MDM, the bias of the model in \textit{Setting 2} minimized. However, in realistic situations we may not know the functional forms of the covariates in the outcome model for matching. 

Compared to PSM and MDM, which are members of the class of equal percent bias reducin (EPBR) and do not guarantee any level of imbalance reduction, CEM is a member of the class of Monotonic Imbalance Bounding (MIB) and is able to guarantee that the imbalance will not be larger than the ex ante user choice~\cite{iacus2012causal}. However, the selection of the number of bins can be challenging.  If a large number of bins are used, the matching performance would be good, but too many samples may be discarded, resulting in inefficient inference. If a small number of bins are used, the number of matched samples may be large, but the matching performance is likely to be poor~\cite{iacus2009cem, iacus2012causal}. As a result,  substantive knowledge of the measurement scale of each variable is needed to decide how much each variable can be coarsened without losing crucial information~\cite{iacus2012causal}. In addition, it is difficult to perform exact matching for data with a large number variables even with very coarse bins, i.e., the curse of dimensionality. Our simulations also show that the performance of the CEM is highly dependent on the choice of the number of bins. In most of our settings, CEM2 had the largest bias and therefore the largest MSE among all models we studied. In contrast, the bias of CEM5 is generally small. However, in some settings (e.g., \textit{Scenario 1} \textit{Setting 1}), CEM5 has a larger SD than CEM2, which leads to a larger MSE. 

\subsubsection{Nonmatching-Based Methods}
The nonmatching methods in general have relatively smaller SDs than the matching based methods as they can utilize information from more samples. 

In many of our settings, the SD and MSE of the basic model LR are very small. However, when the nonlinear effect is present in the outcome model, the bias and the MSE of the LR model can be very large. In addition, the LR is also affected by the violation of the positivity assumption. As shown in our simulation results, the bias of the LR model under \textit{Scenario 2} \textit{Setting 2} is much greater than its bias under \textit{Scenario 1} \textit{Setting 2} because the model dependence issue is exaggerated through extrapolation with positivity violation~\cite{ho2007matching}.

For the inverse probability weighting methods, the performance of the AIPW model in our simulations is always not inferior to, if not better than, the IPW model in terms of bias and MSE, which demonstrates the advantage of the AIPW's doubly robustness in estimation accuracy. However, when the nonlinear effect is present in the outcome model (e.g., \textit{Scenario 1} \textit{Setting 2},  \textit{Scenario 2} \textit{Setting 2}), the AIPW model has a relatively large bias because it does not provide a good estimate of the treatment assignment or the outcome. 

The bias of AIPW\_SL is much smaller than the AIPW when the nonlinear effect is present and comparable to the AIPW in other settings. Attributed to the doubly robustness and SuperLearner's flexibility, the AIPW\_SL always presents a small bias in all of our settings in \textit{Scenarios 1} and \textit{2}. In addition, the SD of AIPW\_SL is also very small for all settings in \textit{Scenarios 1} and \textit{2}. Therefore, the MSE of AIPW\_SL is always one of the smallest, if not the smallest, of all settings for \textit{Scenarios 1} and \textit{2}. Another doubly robust estimator with the SuperLearner, TMLE\_SL, has similar bias/SD/MSE as the AIPW\_SL in \textit{Scenario 1}, but higher SD and MSE in \textit{Scenario 2}. 

\begin{figure}
    \centering
    \includegraphics[width=1\linewidth]{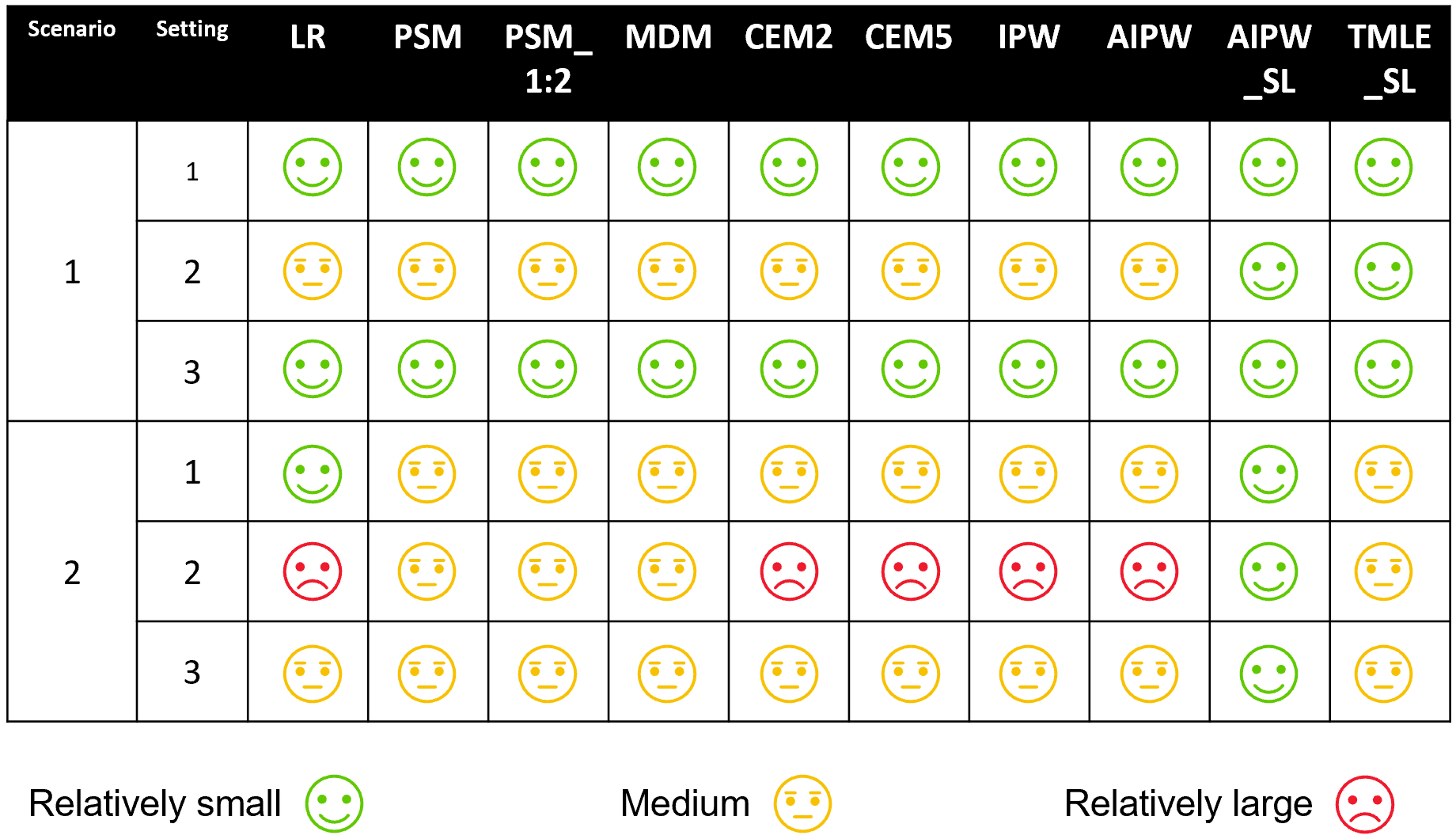}
    \caption{Summary of MSE based on simulations under \textit{Scenarios 1} and \textit{2}. We evaluate models by comparing their relative performance to other models and scenario-setting combinations; therefore, identical icons do not ensure equal performance.}
    \label{fig:mse}
\end{figure}

A summary of MSE based on simulations under \textit{Scenarios 1} and \textit{2} is provided in Figure~\ref{fig:mse}.

Under \textit{Scenario 3} where an important confounder is not considered in estimating ATT, a substantial bias is introduced in all models. Therefore, the bias due to the violation of unconfoundedness cannot be mitigated by model selection. Although the unconfoundedness assumption cannot be formally examined, sensitivity analyses have been developed to assess whether a significant result is sensitive to potential unmeasured confounding.  For example,  a popular sensitivity analysis is based on the E-value measuring the minimum strength of association that an unmeasured confounder would need to have with both the treatment and outcome to explain away the observed causal effect~\cite{vanderweele2017sensitivity}. It is important to note, however, that the application of the E-value is constrained by certain limitations, and its interpretation should be approached with caution~\cite{blum2020use}.

\subsection{Type I error control}
In terms of type I error control, the PSM, PSM\_1:2, and MDM models perform best for all settings in \textit{Scenarios 1} and \textit{2} because of their small biases and accurate estimation of SDs.

In certain scenario-setting combinations, LR, CEM2, CEM5, IPW, and AIPW exhibit inflated type I errors, primarily attributable to biases in their estimation.

The adequacy of type I error control in TMLE\_SL is contingent on the fulfillment of the positivity assumption. Our simulations reveal that when this assumption is breached, TMLE\_SL tends to underestimate the SD, resulting in severe type I error inflation. This result echoes previous findings in the literature that the use of TMLE in estimating the average treatment effect (ATE) produces type I error inflation~\cite{leger2022causal}. The AIPW\_SL encounters a similar issue, though its severity is less pronounced compared to the TMLE\_SL.  

\begin{figure}
    \centering
    \includegraphics[width=1\linewidth]{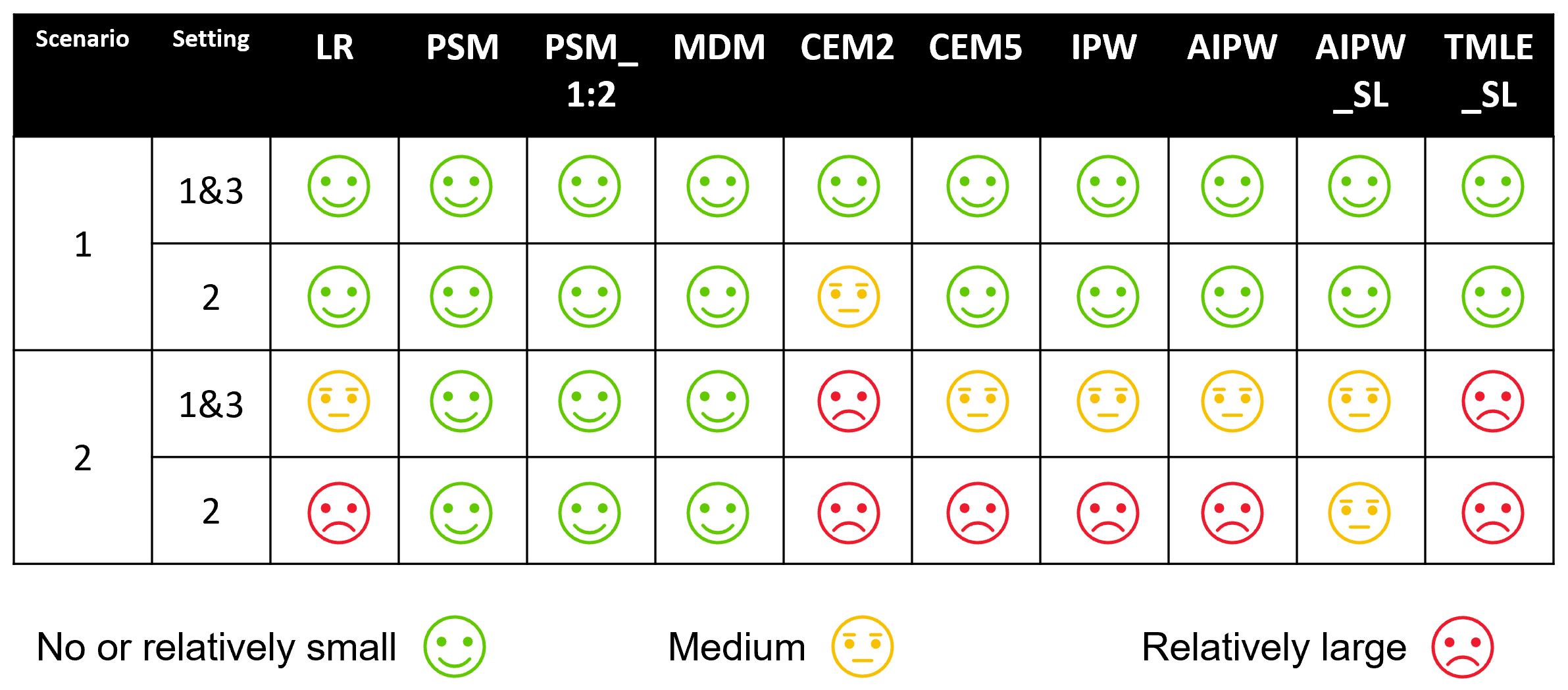}
    \caption{Summary of type I error rates based on simulations under \textit{Scenarios 1} and \textit{2}. We evaluate models by comparing their relative performance to other models and scenario-setting combinations; therefore, identical icons do not ensure equal performance. }
    \label{fig:type1err}
\end{figure}

A summary of type I error rates based on simulations under \textit{Scenarios 1} and \textit{2} is provided in Figure~\ref{fig:type1err}.

\subsection{Scope and limitations of simulation findings}
We would like to emphasize that the primary purpose of our simulations is to identify scenarios and settings that reveal the limitations of the causal inference methods, rather than to provide a comprehensive evaluation under all possible conditions, which is not feasible. As a result, the comparison results for performance of estimation and type I error control may not directly generalize to other simulation settings. When considering specific methods, we recommend that trialists perform thorough evaluations tailored to their particular needs prior to implementation.

\subsection{Considerations on the design and selecting analytic methods for single-arm trials with external controls}
This study compared the performance of various methods for estimating the ATT in single-arm clinical trials with external RWD controls. Through simulations under different identifiability assumptions, we evaluated metrics of 10 causal inference methods, including matching-based and nonmatching-based approaches. We found that the doubly robust methods generally outperformed other approaches in terms of bias. However, no single method consistently outperformed others in terms of MSE. Additionally, the identifiability assumptions played a critical role in the models' performance, with violations of the positivity and unconfoundedness assumptions leading to potential inflation of type I errors and biases. Among the matching-based methods, PSM, PSM\_1:2, and MDM showed robustness under various settings, but their performance was affected by the violation of the positivity assumption and the interaction between the treatment variable and the covariate. By contrast, nonmatching methods in general had smaller SDs than matching-based methods due to their ability to utilize more data. AIPW\_SL  and TMLE\_SL exhibited advantages in terms of bias and MSE due to their double robustness and flexibility, but they tended to underestimate the SD and experienced type I error inflation with violated positivity assumption. 

In light of these findings, it is essential to recognize the limitations of causal inference methods for single-arm trials with external controls, especially when any identifiability assumption can be violated. Violations of these assumptions can lead to biased estimates of treatment effects and an increased risk of type I errors. Therefore, it is crucial to have careful considerations of these factors in both the design and analysis phases of such trials. Accordingly, we would like to recommend the following procedures for planning the single-arm trial with an external control.
\begin{enumerate}
\item \textbf{Determine the causal treatment effect of interest:} To address pertinent clinical questions, it is essential to select the appropriate causal treatment effect of interest. Specifically, when seeking to understand the treatment effect solely in treated patients, attention should be directed towards ATT.
\item \textbf{Consider the consistency assumption in the design phase:}  The consistency assumption requires the treatment be sufficiently well-defined. Therefore, during the design stages, the clinical experts should scrutinize the definition of the treatment to ensure that it is adequately well-defined. Additionally, it is important to ensure that the data to be collected contain values that align with the well-defined treatment.
\item \textbf{Consider the positivity assumption in the design phase:} The positivity assumption cannot be examined at the trial design stage. However, measures should be used to make this assumption as valid as possible. If historical trial patients are selected for inclusion in the external control arm (ECA), the planned single-arm trial and historical trials should share similar inclusion/exclusion criteria and other design elements. For other sources of RWD to be included in an ECA, e.g., electronic medical records, the eligibility criteria for the single-arm and the ECA should also be made to avoid theoretical violation of the positivity assumption. An alternative and preferred way to fulfill this purpose is to use prospectively-collected RWD rather than retrospectively-collected. In addition, if possible, prior study information can be used to evaluate the positivity assumption, e.g., plotting the PS distribution based on historical studies with the same treatment.
\item \textbf{Consider the unconfoundedness assumption in the design phase:} The unconfoundedness assumption can be met if all confounders are identified and controlled. However, if an important confounding factor is not taken into account, the analysis results can be highly biased. Every effort should be made to identify all important confounders through the literature and expert knowledge.   In addition, it should be ensured that the information from all important confounders is measured in the planned single-arm trial and is available in the RWD used for the ECA. Furthermore, sensitivity analyses such as the E-value or other tipping point analyses should be pre-planned during the design stage. 
\item \textbf{Select appropriate analytic models for the primary and sensitivity analyses:} One of the most important requirements for the primary analysis method is to ensure the control of the type I error. After the type I error control is taken into account, the accuracy of the causal effect estimation is the next important consideration. For trials in which the positive assumption cannot be validly justified in the design phase, we suggest using methods that can largely control the type I error even if the positive assumption does not hold, e.g., the PSM or MDM model. When the expected prevalence rate of treatment is low, then 1 to r matching with r $\geq$ 2 should be considered. For experiments where the positivity assumption can be validly justified, a doubly robust approach such as TMLE\_SL or AIPW\_SL may be recommended as the primary analysis method to obtain better estimation precision. 
\end{enumerate}

After the trial is conducted and the trial data are available, one may consider the following procedures to assist in analyses using the causal inference models.

\begin{enumerate}
\item \textbf{Check the performance of the method:} For example, balance diagnostics should always be conducted to evaluate the matching performance of the matching-based method~\cite{zhang2019balance}.
\item \textbf{Evaluate the positivity assumption:} The plot of the PS distribution should be plotted to evaluate the positivity assumption.  If the positivity assumption appears to be clearly violated, then results based on methods that are strongly influenced by the violation of this assumption, such as doubly robust methods with the SuperLearner, may not be reliable and should be interpreted with caution.
\item  \textbf{Conduct sensitivity analyses for the unconfoundedness assumption:} Although the unconfoundedness assumption in general cannot be tested directly, sensitivity analyses should be performed to assess whether the primary analysis results are robust against the unmeasured  confounder. 
\end{enumerate}

\section*{Acknowledgements}
The authors express their gratitude to Dr. Xiaoyu Cai, Dr. Yuan-Li Shen, and Dr. Hana Lee for their valuable and constructive input and participation in discussions during the planning and development of this research work as well as editorial support that greatly enhanced the presentation of this manuscript. 

\section*{Disclaimer} 
The contents, views or opinions expressed in this publication or presentation are those of the authors and do not necessarily reflect official policy or position of the U.S. Food and Drug Administration. Mention of trade names, commercial products, or organizations does not imply endorsement by the U.S. Government.

\section*{Supplementary materials}
Supplementary materials are available online with this paper.

\section*{Funding}
This work was supported by the ORISE Research Program of the U.S. Food and Drug Administration.

\printbibliography

\end{document}